# Possible magnetic phase separation in Ru doped $La_{0.67}Ca_{0.33}MnO_3$

L. Seetha Lakshmi[1], V. Sridharan[1@], D.V. Natarajan[1], Sharat Chandra[1],
V. Sankara Sastry[1], T.S. Radhakrishnan[1], Ponn Pandian,[2] Justine[2] and A. Narayanasamy[2].
[1]Materials Science Division, Indira Gandhi Centre for Atomic Research, Kalpakkam, Tamil Nadu 603 102, India
[2]School of Physics, University of Madras, Guindy Campus, Chennai, Tamil Nadu 600 025, India.

**Abstract:**

X-ray diffraction, resistivity, ac susceptibility and magnetization studies on $La_{0.67}Ca_{0.33}Mn_{1-x}Ru_xO_3$ ($0 \leq x < 0.1$) were carried out. A significant increase in the lattice parameters indicated the presence of mixed valance state of Ru: $Ru^{3+}$ and $Ru^{4+}$. The resistivity of the doped compounds exhibited two features: a broad maximum and a relatively sharp peak. While a para to ferromagnetic transition could be observed for the latter peak, no magnetic signal either in ac susceptibility or in magnetization measurements could be observed for the broad maximum. The magnetic moment decreases non linearly from 3.55 to 3 $\mu_B$ over the Ru composition from 0 to 8.5 at.%. Based on the results of the present studies and on existing literature on the Mn-site substituted systems, we argue that a magnetic phase separation occurs in the Ru doped system. While the sharp peak in the resistivity corresponds to $Ru^{4+}$ enriched region with a ferromagnetic coupling with neighboring Mn ions, the broad peak corresponds to a $Ru^{3+}$ rich regions, with an antiferromagnetic coupling with neighboring Mn ions.

--------------------------------------------
@ Author for correspondence
e-mail: sridh61@hotmail.com; varadu@igcar.ernet.in

## I. Introduction.

The colossal magnetoresistive manganites of the type $REMnO_3$ (RE is tri and/or divalent rare earth ion) continue to attract attention as they exhibit a plethora of phases rich in physical properties: ferromagnetic metal, anti-ferromagnetic insulator, ferromagnetic insulator, cluster glass, charge-ordered and orbital-ordered ground states. Collapse of the manganites into any one of these states is accomplished by altering the subtle balance between the charge, spin, lattice and orbital interaction [1-3]. RE-site substitution, offering one such possibility, has brought out the importance of the carrier density (*n*), Mn-O bond length ($d_{Mn-O}$) and Mn-O-Mn bond angle (<Mn-O-Mn>) in controlling the physical properties of the manganites [3-8]. For a given value of *n*, the bandwidth could be fine tuned to either wide ((La,Sr)-Mn-O system), narrow ((Pr,Ca)-Mn-O system) or intermediate ((La,Ca)-Mn-O system) bandwidth through the variations in values of $d_{Mn-O}$ and <Mn-O-Mn> [3-8]. Though the ferromagnetic metallic (FM-M) state is understood in terms of a double exchange mechanism [9-11], the mechanism is not able to explain the resistivity behavior in the paramagnetic insulating (PM-I) region, especially for the intermediate bandwidth systems [12]. The need to incorporate polarons, due to the presence of Jahn-Teller ion $Mn^{3+}$ with a strong electron-phonon coupling, has been emphasized by Millis *et. al.* for explaining the resistivity variation with temperature [12-14].

The type of magnetic interaction between the neighboring magnetic moments, an important aspect of double exchange interaction, is not addressed by the RE-site substituted studies. In this context, Mn-site substitution provides scope for manipulating the local magnetic coupling between the magnetic moments of the substituents and Mn-ions by suitable substitution of the paramagnetic ion. Numerous investigations have been carried out on Mn-site substitution [15-22]. All the substituents are reported to decrease the transition temperature but to different extent. The decrease in the transition temperature has been attributed by the previous workers to ‘weakening’ of the double exchange interaction upon substitution. In our previous studies on Mn-site substitution with paramagnetic and diamagnetic ions [23], the variation in the value of transition temperature suppression rate with concentration, $dT_C/dx$, was rationalized in terms of local structural disorder and the nature of local magnetic coupling between the magnetic moments of the substituents and Mn ions. From the value of $dT_C/dx$ for the Ru-substituted system, we had speculated on the nature of magnetic coupling between the Ru and Mn ions as being ferromagnetic in nature.

Importance of magnetic interaction by substituting various para and diamagnetic ions in the charge ordered Nd-Sr-Mn-O [24] and Pr-Sr-Mn-O [25] systems (referred as CO-I in this work). More dramatic effects of Mn-site substitution on the CO-I state of the $Pr_{0.5}Ca_{0.5}MnO_3$ were brought out by Hebert *et. al.* [26]. Although the diamagnetic substitutions (e.g. $Ti^{4+}$, $Zr^{4+}$, $Ga^{3+}$, $Sn^{4+}$) melt the charge ordered state of $Pr_{0.5}Ca_{0.5}MnO_3$ to result in a spin glass state, the system continues to be insulators. On the other hand paramagnetic substitutions (e.g. $Ru^{4+/5+}$, $Co^{2+}$, $Ni^{2+}$, $Cr^{3+}$), baring $Fe^{3+}$ which behaves like a paramagnetic ions in its effect on the CO-I state, not only melts the charge ordered state but also rendered the ground state to be ferromagnetic metal. Among these paramagnetic substitutions, metal to insulator (MI) transition temperature of the Ru substituted system was reported to be as high as 240 K while for others it was typically less than 150 K. Thus, Ru substitution assumes an important role in elucidating the role of magnetic interaction in manganites.

In this paper, we report the effect of Ru substitution on the magnetic and transport properties of the archetypical colossal magnetoresistive compound, viz., $La_{0.67}Ca_{0.33}MnO_3$, (referred as CMR in this work). From the increase in the lattice parameters, it is shown that Ru has a mixed valance state of $3^+$ and $4^+$ and not $4^+$ and $5^+$ as reported in earlier works. The Ru substitution results in two peaks in the resistivity curve: a sharp peak followed by a broad maximum at a still lower temperatures with a suppression rate of ~2.3 and 17 K/at.% respectively. From the resisitivity and magnetization studies on $La_{0.67}Ca_{0.33}Mn_{1-x}Ru_xO_3$ (x=0, 0.01, 0.03, 0.05, 0.07 & 0.085) it is reasoned that Ru substitution results in a magnetic phase separation of a weak ferromagnetic phase within a ferromagnetic matrix. We also argue that such a phase separation is generic nature of the paramagnetic substituted system with FM/M as the ground sate, irrespective of the nature of the parent compound i.e. CMR or CO/I.

**II. Experimental.**

The compounds were synthesized by standard solid-state reaction. The stoichiometric mixture of the $La_2O_3$, $CaCO_3$, $MnO_2$ and $RuO_2$ were heat treated in the temperature range 1000 to 1500$^oC$ in flowing oxygen atmosphere with three intermediate grindings followed by pelletization. Final sintering of the samples in flowing oxygen was carried out in single batch to ensure that the samples are subjected to identical sintering conditions. $La_2O_3$ was pre-calcined at 800$^oC$ for 12 hours to remove moisture and was subsequently weighed. The density of the sintered pellets was determined by the standard Archimedes principle, using methanol as the liquid. The room temperature powder X-ray diffraction (XRD) pattern in the reflection geometry with high statistics ($10^5$ counts over a dwell time of 20 sec at the 100% peak) was recorded using $Cu_{K\alpha}$ radiation (STOE, Germany). The lattice parameters and the fractional co-ordinates were determined using RIETAN refinement programme [27]. In estimating the fractional co-ordinates, the site occupancies of the atoms were fixed in the ratio of nominal composition of the compounds. The resistivity in the Van der Pauw geometry was measured in the temperature range 4.2 to 300K using silver paint for the electrical contacts. AC susceptibility was measured in the temperature range 4.2 to 300K using a home built ac susceptometer under an average field of 25 $\mu$T at a frequency of 947 Hz. The magnetic transition temperatures ($T_C$) were determined by a tangent method and correspond to the onset of the $\chi$' signal. The hysteresis loops were recorded at 80K using a vibrating sample magnetometer with a maximum sweep field of 0.7T. The samples were degaussed before recording the hysteresis loop. The grain size was examined using JEOL Scanning Electron Microscope (SEM) (Model: JSM 5410).

**III. Results and discussion.**

Figure 1(a) shows the room temperature powder X-ray diffraction pattern of $La_{0.67}Ca_{0.33}Mn_{1-x}Ru_xO_3$ for x = 0,0.01,0.03,0.05,0.07,0.085 and 0.1. In all these compounds, other than for x = 0.1, no impurity phases could be detected. For x = 0.1 sample, an impurity peak could be observed at $2\theta = 43.3^o$ (inset) with a relative intensity of 3% and this composition will not be included in our consideration. All the patterns could be indexed to orthorhombic symmetry of space group P*nma*. The variation of the lattice parameter *a*, *b* and *c* with Ru concentration is tabulated in Table I and are shown in Fig 1(b). While the lattice parameters *a* and *c* systematically increase with Ru concentration (~0.25% for x=0.085), marginal increase (~0.08% for x=0.085%) was observed for the lattice parameter *b*. From the fractional co-ordinates, the average Mn-O bond length ($d_{Mn-O}$) and O-Mn-O bond angle (<O-Mn-O>) were estimated and are also given in Table. 1. However, we could not observe any systematic changes in the values of bond length and bond angle with Ru composition, though the lattice parameters increase with Ru composition. This is basically due to the difficulties in obtaining a reliable estimate of the fractional co-ordinates of the oxygen from the powder X-ray diffraction pattern, in presence of other strong scattering species like La, Ca and Mn.

The variation of resistivity ($\rho$) and ac susceptibility with temperature (in-phase component $\chi$' alone is shown) for the compounds is shown in Fig. 2 and 3 respectively. The undoped sample (x=0) exhibits a metal ($d\rho/dT > 0$) to insulator ($d\rho/dT < 0$) transition marked by the presence of a peak in the resistivity curve (Fig. 2). Around this temperature, a paramagnetic to ferromagnetic transition is also observed (Fig. 3), a characteristic feature of the colossal magnetoresistive manganites [1-3]. In contrast to this, the doped samples exhibit two MI transitions: a relatively sharper peak followed by a broad maximum at a still lower temperature. While the high temperature peak is denoted as $T_{P1}(\rho)$, the broad maximum is denoted by $T_{P2}(\rho)$. At about $T_{P1}$, the doped samples exhibit a ferromagnetic to paramagnetic transition (Fig. 3). It is pertinent to note here that no additional signal in the $\chi$' or magnetization (not shown) corresponding to the broad resistivity maximum is observed. The transition temperatures $T_{P1}(\rho)$, $T_{P2}(\rho)$ and $T_C$ are tabulated in Table II. For the lower concentration of the substituent ($0< x \leq 0.03$), $T_{P2}(\rho)$ could not be estimated reliably as it is rather weak and strongly overlaps with the other one. It is seen that, both the transition temperatures $T_{P1}(\rho)$ and $T_C$ are lowered with substitution (Figure 4) at a rate of $\approx$ 2.3 K/at.%, comparable to that of Ru doped La-Sr-Mn-O system [28]. On the other hand, the broad maximum shifts to lower temperatures with increasing Ru concentration at a rate of 17 K/at.%, which is comparable to that for Fe or Ga substituted (La,Ca)-Mn-O systems [23].

In Figure 5, the hysteresis loops of the compounds measured at 80 K are compared. The saturation of the magnetization for all the compounds was realized with an applied field $H_a \geq 0.6$ T. As the temperature of the hysteresis loop measurement is typically less than 0.3 $T_C$, the saturation magnetization measured at 80 K is taken to correspond to $M_S(0)$. The magnetic moments $\mu(x)$ were estimated from the saturation magnetization values and its variation with the Ru concentration is shown in Fig. 6. The magnetic moment of the x=0 compound is determined to be 3.55 $\mu_B$, which is very close to the expected value of 3.67 $\mu_B$. It is pertinent to note here in the first place that the variation of $\mu(x)$ is *not linear* over the entire Ru concentration and that $\mu(x)$ drops from the value of 3.55 $\mu_B$ for the undoped compound to ~3 $\mu_B$ for the x=0.085 compound.

The valance state of Ru in these compounds is a subject of debate. Among the possible valance states, the states pertinent to the present study are $3^+$, $4^+$ and $5^+$ with ionic radii of 0.68, 0.62 and 0.565 Å respectively for co-ordination number of six [29]. As Ru doping melts the charge ordered state of $Pr_{0.5}Ca_{0.5}MnO_3$ and sustains the ferromagnetic metallic state [26], Hebert *et. al.* have proposed the presence of $Ru^{4+}$ and $Ru^{5+}$ (isoelectronic to the double exchange couple, $Mn^{3+}$ and $Mn^{4+}$) in these compounds. Such a combination has a weighted ionic radius of $\langle r \rangle_w = 0.602$ Å, which is comparable to the weighted radius of $Mn^{3+}$ and $Mn^{4+}$, namely 0.607 Å. On the other hand, the observed appreciable increase in the value of the lattice parameters *a*, *b* and *c* (Fig. 1(b)) in the present study clearly indicates that the weighted ionic radius of ruthenium should be larger than that of manganese. Though $Ru^{3+}$, (isoelectronic to $Fe^{3+}$) has a larger ionic radius (0.68 Å) compared to that of $Mn^{3+}$ and $Mn^{4+}$ pair, inferences from other measurements as will be shown subsequently, do not favour the presence of $Ru^{3+}$ alone. Hence, Ru should have a mixed valance of $Ru^{3+}$ and $Ru^{4+}$ having $\langle r \rangle_w = 0.66$ Å which could lead to the observed increase in the lattice parameters. Also, X-ray Photoemission Spectroscopic study on Ru substituted (La,Sr)-Mn-O system of similar composition has established the presence of $Ru^{3+}$ and $Ru^{4+}$ couple [28], supporting our inference.

As mentioned earlier, the variation of $\mu$ with the composition is not linear in the whole composition range. However, the variation is more or less linear in the composition range $0<x\leq 0.05$. We have attempted to fit this range with weighted average magnetic moment for various combination of Ru valance states $Ru^{3+}$, $Ru^{4+}$ and $Ru^{5+}$ invoking either an exclusive or a random substitution of $Mn^{3+}$ or $Mn^{4+}$. Obviously, for any such combination of the valence states, an exclusive ferromagnetic interaction between the magnetic moments of Ru and Mn ions cannot explain the *decrease* in value of $\mu(x)$. Though an exclusive antiferromagnetic coupling between Ru and Mn ions would result in a decrease of $\mu$ with x, still the observed reduction is larger. Such deviations, which were also reported for the Cu substituted

La-Sr-Mn-O compounds [30], clearly indicate that the sample is magnetically inhomogeneous. As will be shown subsequently, the picture is somewhat complicated due to magnetic phase separation occurring in this system.

In addition to and somewhat correlated to the broad maximum in the resistivity curve is the observation of an up-turn in the resistivity below 35 K. In passing, we also wish to mention the following features observed in the resistivity curves (Fig. 2): **a.)** The resistivity of the doped samples for $0< x < 0.05$ are lower than that of undoped one over the entire temperature range of the measurement and **b.)** Among the substituted compounds, the residual resistivity ($\rho_o$) increases with Ru concentration (Figure 6). Both the features (presence of broad maximum and an up-turn in the resistivity), have been reported for the polycrystalline samples and manifest dominantly in the samples with smaller grain sizes [31-33]. These features are conspicuously absent in the case of single crystals and high quality thin films. The up-turn in the resistivity of the polycrystalline samples is shown to arise from the inter-grain tunneling of the spin-polarized ($e_g$) conduction electrons [34,35]. On the other hand, the broad maximum in the metallic region ($T < T_P$) corresponds to grains substantially smaller in size and rich in defect structures (due to enhanced grain boundary region), leading to a lowered transition temperature with a distribution in $T_C$. It is pertinent to consider the origin of the hump seen in the present study, namely the reduction in the grain size with doping in the light of the above reasoning.

The density of samples, using the Archimedes method, is within the range 92 to 97% of the theoretical value and with no systematic variation in the density with x. In Fig. 7, the SEM picture of the undoped and x=0.01,0.05 and 0.085 are shown as representative. It is seen that the grain size in the case of undoped compound is ~ 20 μm and contains regions of poor inter-grain connectivity (Fig. 7(a)). Such a poor connectivity between the grains can lead to the observed up-turn in the resistivity. In the case of doped samples, an overall improvement in the inter-grain connectivity (Fig. 7(b-d)) is observed. This results in lowering the resistivity of the doped sample and decreases the up-turn tendency, especially for $0< x \leq 0.05$. Though a slight reduction in the grain size can be seen with doping (Fig. 7 (c & d)), the average grain size is not small enough, in our opinion, to effect a broad maximum of the type reported in Ref. 31.

Ours is not the only study to report the broad maximum in the metallic phase *in the doped* systems. Resistivity curves of the Cr, Ni, Co and Cu (all paramagnetic ions with unfilled $e_g$ orbital) substituted colossal magnetoresistive systems exhibit a hump in the metallic phase [20,36,37]. Such a maximum is also observed in Ru, Cr, Ni, Co and Ir substituted charge ordered $Pr_{0.5}Ca_{0.5}MnO_3$ [26]. This leads to an interesting observation that doping with *certain paramagnetic ions* leading to a FM-M ground sate, results in a broad maximum in the resistivity curve irrespective of ground state of the parent compound. It thus appears that the additional hump is an intrinsic and generic feature of the paramagnetic (Ru, Co, Ni, Cr and Cu) substituted systems with FM-M ground state, except for Fe. $Fe^{3+}$ ion, although carrying a magnetic moment of S=5/2 has been shown to couple antiferromagnetically [38] with the neighbouring Mn ions and has a half filled electronic ($3t_{2g}$, $2e_g$) structure. These two factors exclude $Fe^{3+}$ ion from participating in the DE interaction. Hence, effect of $Fe^{3+}$ substitution is quite the same as that of diamagnetic ion (such as $Zr^{4+}$, $Ti^{3+}$ and $Ga^{3+}$) substitutions [ 23 ].

In these substitutional studies, except for Cu [37], no additional magnetic signal that can be associated with a broad maximum in the resistivity has been reported, in agreement with our results. In one of the reports on the Cu substituted system [37], additional signal was discerned in the dM/dT curve and was attributed to two distinct *electronically* phase separated regions. Evidence for a phase separation in the case of Cu substituted system, was obtained from the ESR and La-NMR [39,40] studies carried out across the transition temperature.

Extending this reasoning, we hypothesize that phase separation also occurs in case of other paramagnetic (excluding $Fe^{3+}$) substituted systems leading to hump in the resistivity curve in the metallic region. The reason for the observation of a magnetic signal corresponding to the hump in the resistivity for the Cu substituted system in contrast to others could be that macroscopic (on relative terms) phase separation might have occurred in case of Cu substituted system.

The phase separation envisaged in the present work is different from that of the electronic phase separation occurring in other systems[41-43]. In the case of the latter, antiferromagnetic insulator and ferromagnetic metal coexist at low temperature. The volume fraction of these two phases could easily be altered by a suitable thermodynamic variable such as the magnetic field, pressure or by chemical doping as in the case of $(La_{1-x}Pr_x)_{0.67}Ca_{0.33}MnO_3$ [41]. With increase of the AFM-I phase, say by increasing Pr concentration, the residual resistivity is found to increase *several orders* of magnitude. Whereas, in the present study the residual resistivity increases only by a factor of five. Also, in the case of electronic phase separation, additional peak is not expected as only one of the two phases has a metal to insulator transition. Thus, we conclude that the phase separation encountered in the Ru doped $La_{0.67}Ca_{0.33}MnO_3$ is *not* an electronic phase separation as both the phase have a ferromagnetic metallic ground state. Nonetheless, they still could differ in their magnetic properties. The difference in their magnetic properties, influenced by the strength of the DE interaction, can be rationalized in the following way.

From the increase in the value of the lattice parameters on Ru substitution, presence of $Ru^{3+}$ and $Ru^{4+}$ has been inferred. The $T_{P1}(\rho)$ (and the $T_C$) and $T_{P2}(\rho)$ correspond to metal to insulator transition of $Ru^{4+}$ and $Ru^{3+}$ rich regions in the phase-separated scenario with associated transition temperature suppression rate of 2.3 and ~17 K/at.% respectively. $Ru^{3+}$, with its half filled shell electronic configuration, could not participate in the DE interaction and due to ubiquitous superexchange interaction couples antiferromagnetically with neighboring Mn ions [44]. Such a local AFM coupling would weaken the over all DE interaction strength of the concerned phase which in turn results in lowered transition temperature ($T_{P2}$) as well as a larger suppression rate of ~ 17 K/at.% . Indeed this suppression rate is comparable to that of $Fe^{3+}$ substituted system [23]. On the other hand, $Ru^{4+}$ which is isoelectronic to $Mn^{3+}$ can ferromagnetically couple with neighboring Mn ions and also participate in the DE mediated conduction process leading to a lowered suppression rate of 2.3 K/at.% associated with $T_{P1}(\rho)$ and $T_C$. Progressive lowering of the $T_{P1}$ and $T_{P2}$ with Ru concentration clearly indicates that the corresponding phases are getting progressively enriched in $Ru^{4+}$ and $Ru^{3+}$ respectively. Either the enrichment of $Ru^{3+}$ and/or an increase in the volume fraction of the weaker ferromagnetic phase would have resulted in the observed rise in the value of $\rho_o$ among the substituted system.

To sum up, the substantial increase in the lattice parameter values of $La_{0.67}Ca_{0.33}Mn_{1-x}Ru_xO_3$ indicates the mixed valance state for Ru: $Ru^{3+}$ and $Ru^{4+}$. It undergoes a *magnetic phase separation* consisting of a weaker ferromagnetic phase of microscopic size within a ferromagnetic matrix. The magnetic phase enriched in $Ru^{3+}$ corresponds to the weak ferromagnetic microscopic phase with antiferromagnetic interaction with neighboring Mn ions. The phase enriched with $Ru^{4+}$ establishing ferromagnetic interaction with neighboring Mn ions results in ferromagnetic matrix corresponding to the sharp peak in the resistivity curve. We further argue that other paramagnetic substitutions, baring $Fe^{3+}$, leading to a ferromagnetic ground state would also be expected to under go a phase separation and would exhibit a broad maximum in the resistivity.

**ACKNOWLEGEMENTS**:

We thank Dr. Hariharan, Ms S. Kalavathi and Ms T. Geetha Kumary for extending the low temperature facilities. We thank Dr. G.V.N. Rao, Head, Quality Control, ARC,

Hyderabad for the SEM studies.

## Figure captions

1. Room temperature powder X-ray diffraction pattern of $La_{0.67}Ca_{0.33}Mn_{1-X}Ru_XO_3$ system (x=0,0.01,0.03,0.05,0.07,0.085 and 0.1). Inset shows the presence of the only observable impurity peak. Note the intensity is in plotted in logarithmic scale.
2. Temperature variation of resistivity of $La_{0.67}Ca_{0.33}Mn_{1-X}Ru_XO_3$ system (x=0, 0.01, 0.03, 0.05, 0.07, 0.085 and 0.1) as a function of temperature. Inset shows the resistivity in the low temperature region on the expanded scale for better visualization of the resistivity up-turn effect.
3. Temperature variation of ac susceptibility Vs temperature for $La_{0.67}Ca_{0.33}Mn_{1-X}Ru_XO_3$ system (x=0, 0.01, 0.03, 0.05, 0.07, 0.085 and 0.1) .
4. Variation of the ferromagnetic to paramagnetic transition ($T_C$) and the metal to insulator transitions ($T_{P1}$ and $T_{P2}$) as function of Ru concentration of $La_{0.67}Ca_{0.33}Mn_{1-X}Ru_XO_3$. Refer the text for the details.
5. Hysteresis loop of $La_{0.67}Ca_{0.33}Mn_{1-X}Ru_XO_3$ (x=0,0.01,0.03,0.05,0.07,0.085) at 80K. Inset shows the variation of coercive field $H_C$ with the ruthenium concentration.
6. Variation of magnetic moment and the residual resisitivity with ruthenium concentration.
7. SEM picture of $La_{0.67}Ca_{0.33}Mn_{1-X}Ru_XO_3$ (x=0 [a], 0.01[b], 0.05[c] and 0.085[d] )

( as representatives of the series)

L .Seetha Lakshmi *et al*

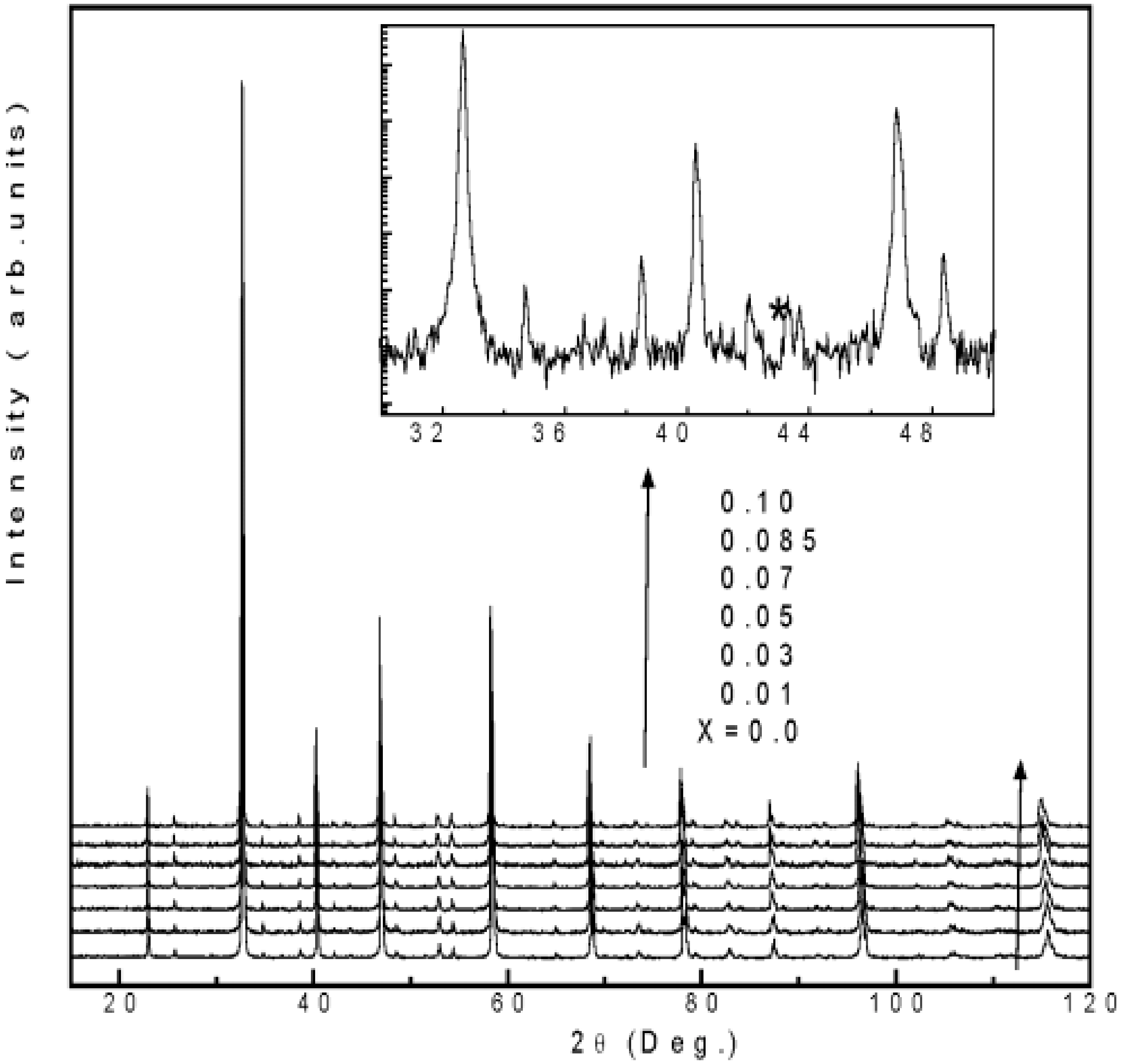

**Figure 1**

L.Seetha Lakshmi *et al.*

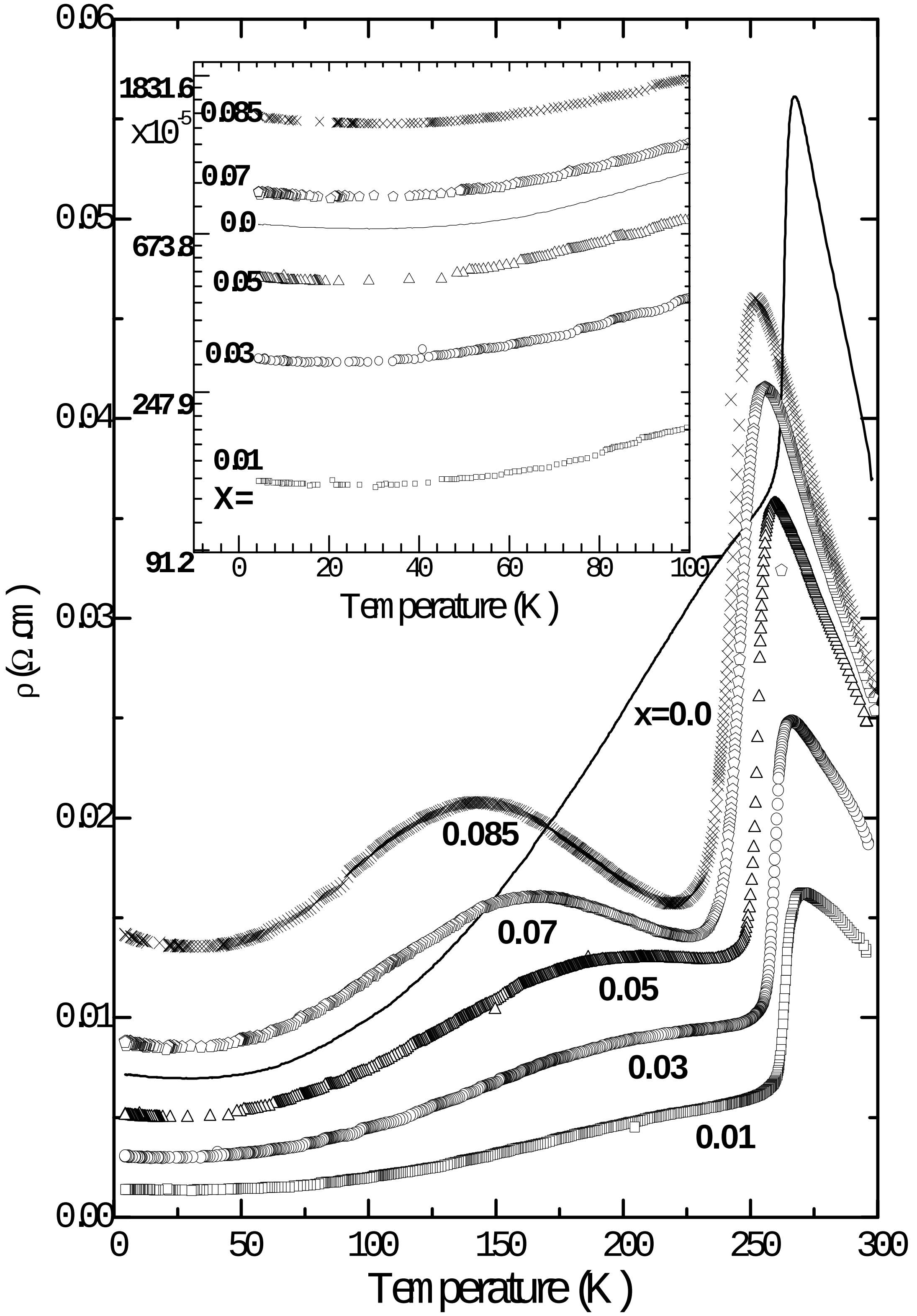

Figure 2

L. Seetha lakshmi *et al*

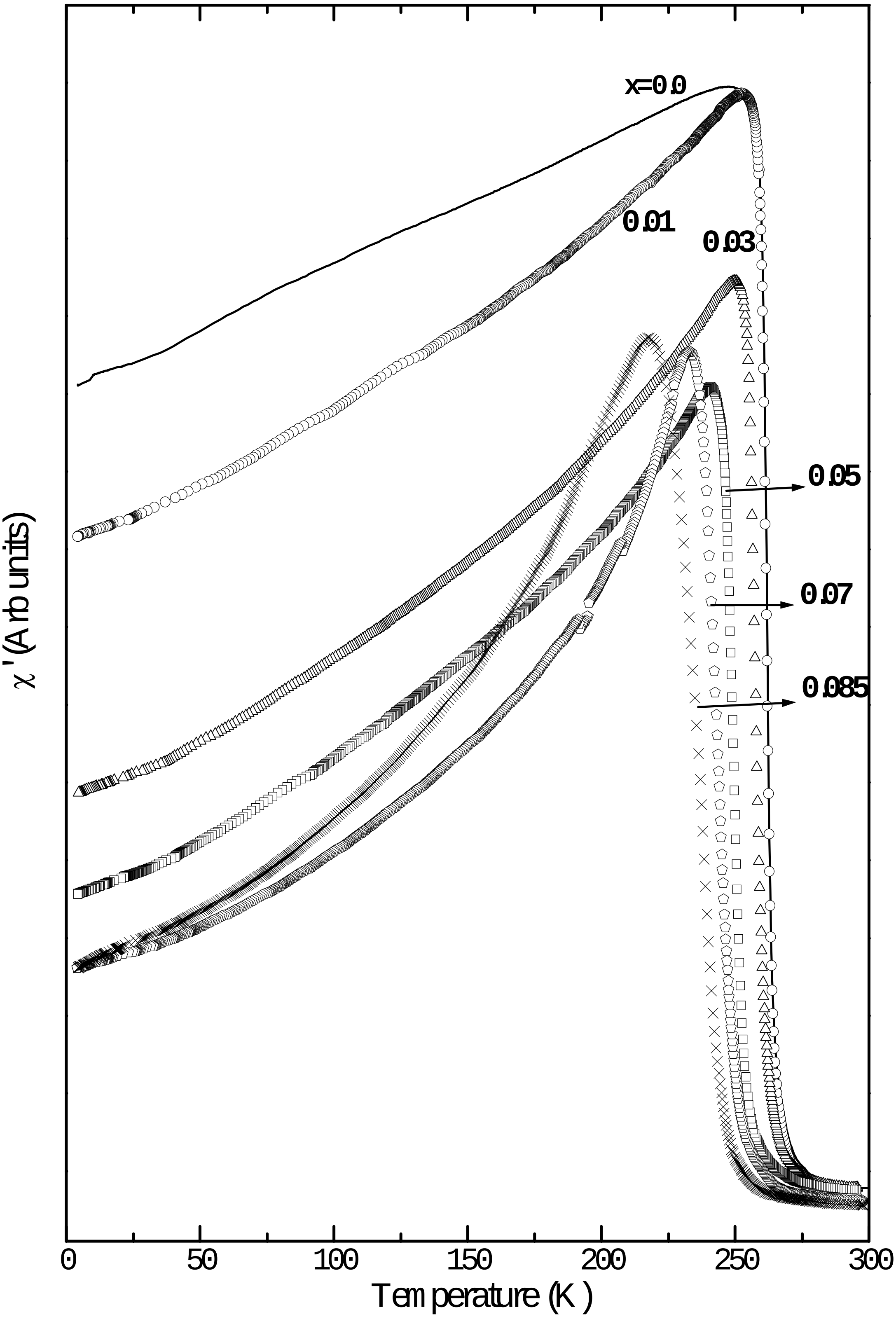

Figure 3

L.Seetha lakshmi *et al*

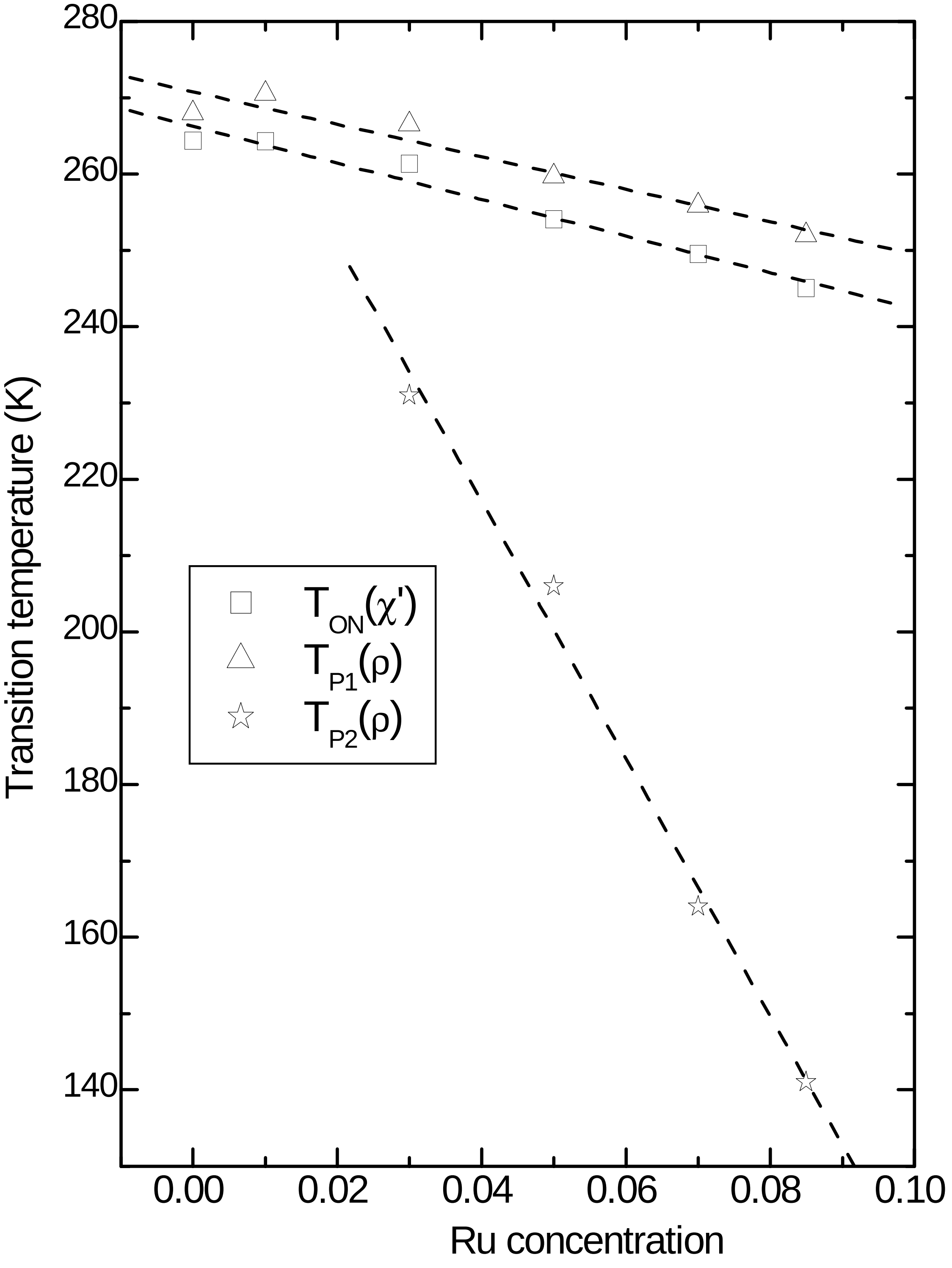

Figure 4

L.Seethalakshmi *et.al.*

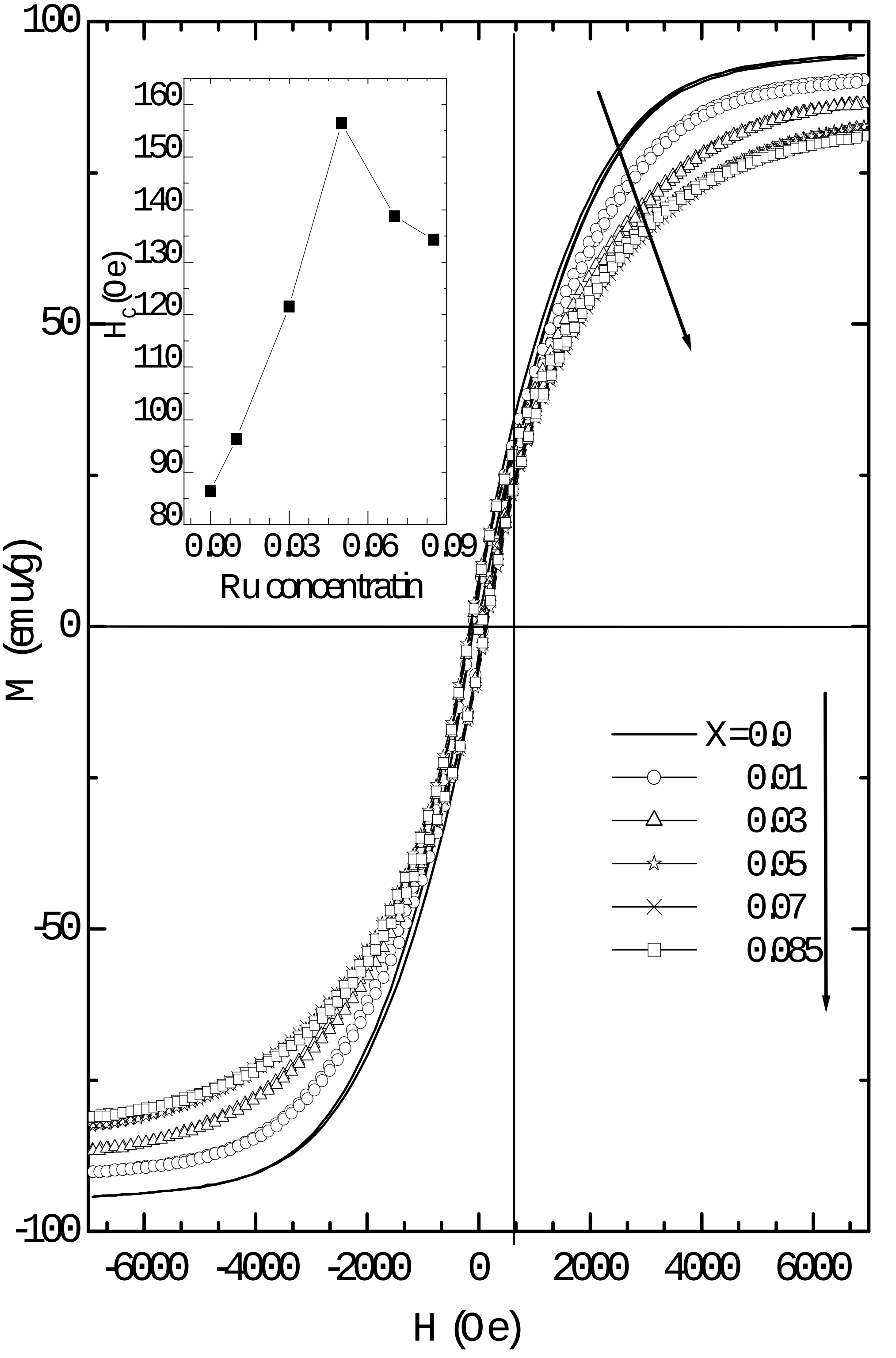

Figure 5

L.Seetha Lakshmi *et.al.*

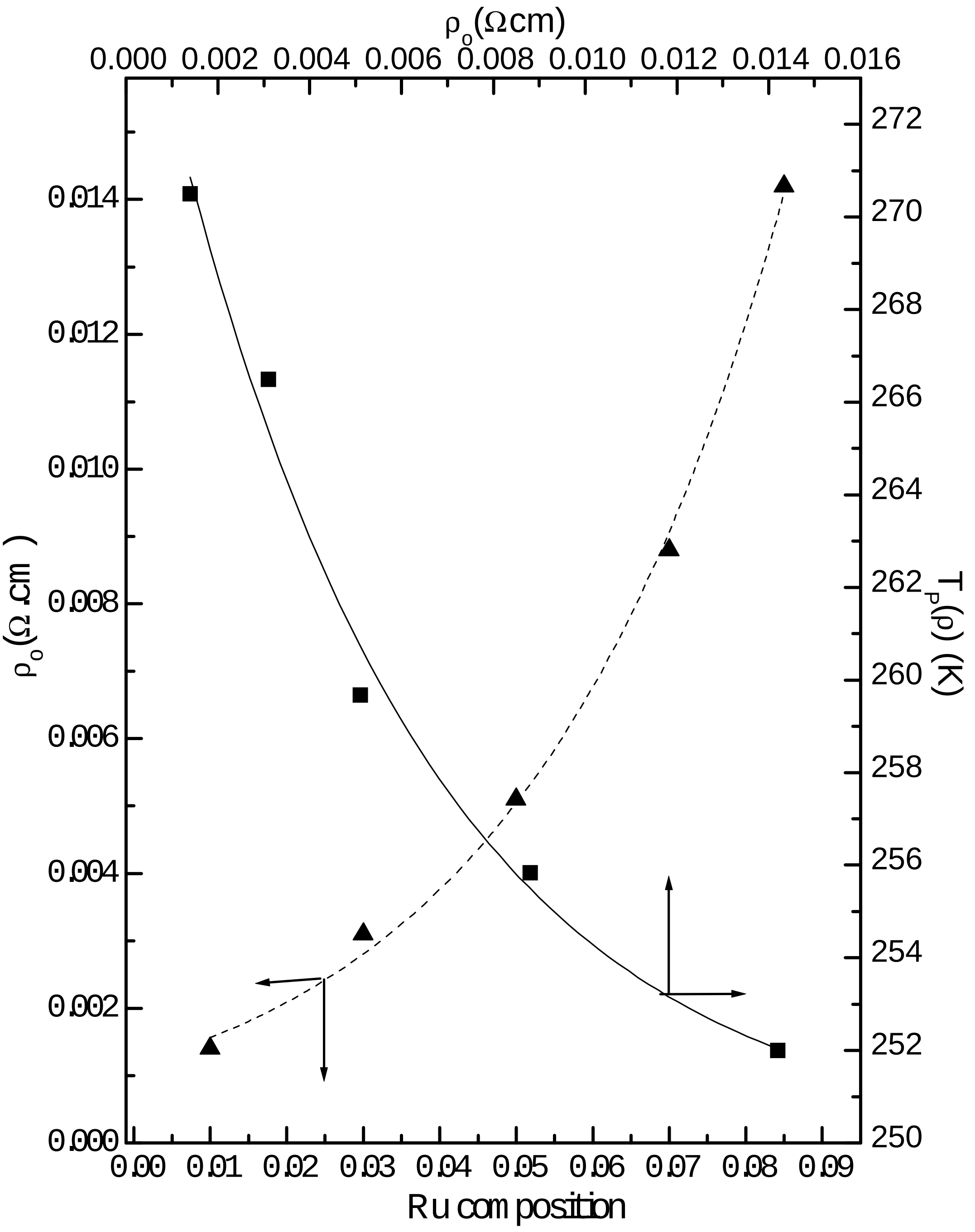

Figure 6

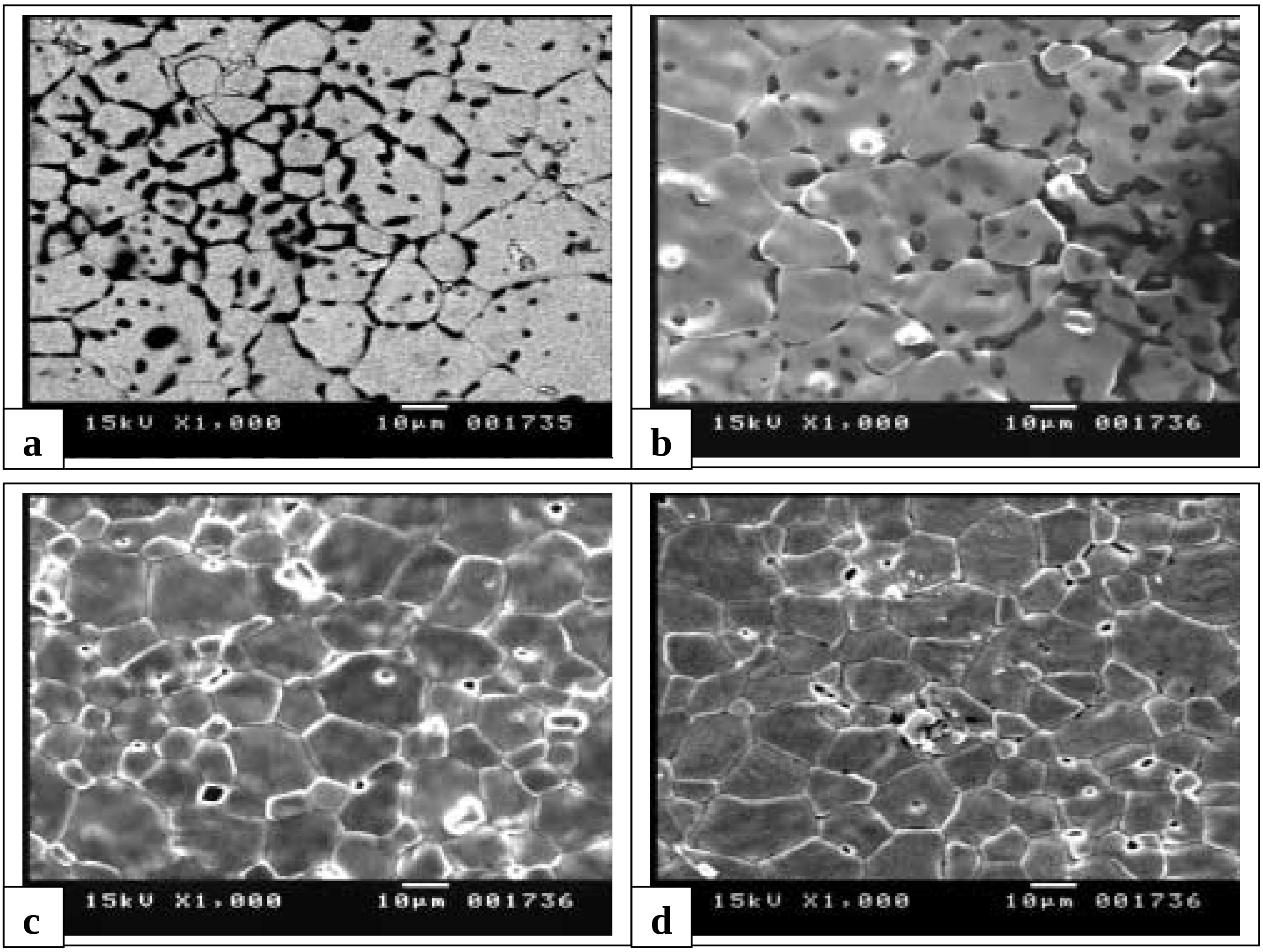
15kV X1,000 10µm 001735
a
15kV X1,000 10µm 001736
b
15kV X1,000 10µm 001736
c
15kV X1,000 10µm 001736
d

Figure 7

Table I: Variation of the lattice parameters, Mn-O bond lengths, $d_{Mn1-O}$ & $d_{Mn-O2}$ (in Å), bond angles Mn-O1-Mn, Mn-O2-Mn bond angel, (in deg.) and the S and $R_{W-P}$ parameters with Ru composition.

| X | a (in Å) | b (in Å) | c (in Å) | $d_{Mn-O1}$ (in Å) | $d_{Mn-O2}$ (in Å) | <Mn-O1-Mn> (in deg.) | <Mn-O2-Mn> (in deg.) | S | $R_{W-P}$ |
|---|---|---|---|---|---|---|---|---|---|
| 0 | 5.4624 | 5.4763 | 7.7201 | 1.9598 | 1.9558 | 160.2 | 162 | 1.13 | 18.07 |
| 0.01 | 5.4624 | 5.4763 | 7.7164 | 1.96 | 1.9558 | 160.2 | 161.9 | 1.13 | 18.07 |
| 0.03 | 5.4659 | 5.4778 | 7.7231 | 1.9624 | 1.9613 | 160.7 | 159.7 | 1.17 | 18.85 |
| 0.05 | 5.4708 | 5.4806 | 7.7294 | 1.9735 | 1.9601 | 157.6 | 160.7 | 1.22 | 12.44 |
| 0.07 | 5.4727 | 5.4797 | 7.7321 | 1.9702 | 1.9577 | 158.7 | 161.8 | 1.13 | 27.15 |
| 0.085 | 5.4759 | 5.4808 | 7.7359 | 1.9736 | 1.9646 | 157.8 | 159.7 | 1.06 | 21.87 |

Table II. Variation the M-I transition temperatures $T_{P1}$ & $T_{P2}$, estimated form the resistivity studies, Tc corresponding to the onset of ac susceptitbility signal, residual resistivityand resistivity at room temperature.

| X | $T_{P1}(\rho)$ in K | $T_{P2}(\rho)$ in K | $T_{ON}(\chi')$ in K | $\rho_o$ in Ω.cm | ρ(290K)in Ω.cm |
|---|---|---|---|---|---|
| 0 | 267.94 | - | 264.34 | 0.00 | |
| 0.01 | 270.5 | - | 264.3 | 0.0014 | 0.0142 |
| 0.03 | 266.5 | 231 | 261.35 | 0.0031 | 0.0206 |
| 0.05 | 259.67 | 206 | 254.04 | 0.0051 | 0.0267 |
| 0.07 | 255.84 | 164 | 249.55 | 0.0088 | 0.0289 |
| 0.085 | 252 | 141 | 245.06 | 0.0142 | 0.0306 |